\documentclass[useAMS,usenatbib]{mn2e}
\usepackage{amsmath}
\usepackage{rotating}

\title[Magnification relations and Einstein crosses]{Magnification relations of quad lenses and applications on Einstein crosses}
\author[Z. Chu et al.]{Zhe Chu$^{1}$\thanks{E-mail:
chuzhe@pmo.ac.cn}, G. L. Li$^{1}$, W. P. Lin$^{2,3}$ and H. X. Pan$^{1}$  \\
$^{1}$Purple Mountain Observatory, Chinese Academy of Sciences, 2 West Beijing Road, Nanjing 210008, China \\
$^{2}$School of Physics and Astronomy, Sun Yat-Sen University, Guangzhou 510275, China \\
$^{3}$Shanghai Astronomical Observatory, Chinese Academy of Sciences, 80 Nandan Road, Shanghai 200030, China}
\begin{document}

\date{Accepted 2015 January 25. Received 2014 November 4; in original form 2014 October 11}

\pagerange{\pageref{firstpage}--\pageref{lastpage}} \pubyear{2014}

\maketitle

\label{firstpage}

\begin{abstract}
In this work, we mainly study the magnification relations of quad lens models for cusp, fold and cross configurations. By dividing and ray-tracing in different image regions, we numerically derive the positions and magnifications of the four images for a point source lying inside of the astroid caustic. Then, based on the magnifications, we calculate the signed cusp and fold relations for the singular isothermal elliptical lenses. The signed fold relation map has positive and negative regions, and the positive region is usually larger than the negative region as has been confirmed before. It can also explain that for many observed fold image pairs, the fluxes of the Fermat minimum images are apt to be larger than those of the saddle images. We define a new quantity \emph{cross relation} $R_{\textrm{cross}}$ which describes the magnification discrepancy between two minimum images and two saddle images. \emph{Distance ratio} $d_{\textrm{sadd}}/d_{\textrm{mini}}$ is also defined as the ratio of the distance of two saddle images to that of two minimum images. We calculate the cross relations and distance ratios for nine observed Einstein crosses. In theory, for most of the quad lens models, the cross relations decrease as the distance ratios increase. In observation, the cross relations of the nine samples do not agree with the quad lens models very well, nevertheless, the cross relations of the nine samples do not give obvious evidence for anomalous flux ratio as the cusp and fold types do. Then, we discuss several reasons for the disagreement, and expect good consistencies for more precise observations and better lens models in the future.
\end{abstract}


\section{Introduction}

Elliptical lens is very important both in theory and observations for modelling triaxial ellipsoid dark matter haloes. For non-singular smooth elliptical lenses, it can produce five images at most for a single point source \citep{bur81,sch92b}. However, the Fermat maximum image located near the lens centre is usually highly demagnified and faint, resulting in four observed images. In the observations of strong gravitational lenses, there is a large proportion of four-image lens systems among all of the multiple lensed quasars. The main reason is that most of the dark matter haloes are triaxial ellipsoid \citep{eva00,jin02}, and their planar projections should correspond to the elliptical lenses \citep{kas93}.

In general, the galaxy lens is not smooth and there is no mass density singularity in galaxy centre. However, to simplify the problem we usually use the analytical smooth lens model with a singular point in the lens centre, i.e., singular isothermal elliptical lens. The tangential critical curve of the SIE lens is an ellipse in the lens plane (image plane), where the Jacobian matrices vanish and the magnifications are infinite. The critical curve divides the lens plane into image regions of positive and negative parities. The caustics are the corresponding curves obtained by mapping the critical curves into source plane via lens equation $\boldsymbol{\beta}=\boldsymbol{\theta}-\boldsymbol{\alpha}$. The radial critical curve of the SIE lens degenerates into a point in the lens centre, and corresponds to the pseudo-caustic \citep{eva98}. The tangential caustic of a typical elliptical lens commonly comprises four cusps and four folds.

The cusp and fold relations are local magnification relations. As shown in the top panel of Fig. 1, if a point source moves to the cusp from the inner side of the tangential caustic, three of the images will merge together near the critical curve. The three close images have an asymptotic magnification relation \citep{bla86,sch92a,sch92b,mao92,chu15}
\begin{equation}
R_{\textrm{cusp}}=\frac{S_{\textrm{cusp}}}{S_{|\textrm{cusp}|}}= \frac{\mu_{\textrm{A}}+\mu_{\textrm{B}}+\mu_{\textrm{C}}}{|\mu_{\textrm{A}}|+|\mu_{\textrm{B}}|+|\mu_{\textrm{C}}|} ,
\end{equation}
where $\mu$ are the signed magnifications of the triple images A, B and C. Here, $S_{\textrm{cusp}}$ and $S_{|\textrm{cusp}|}$ are cusp summation and cusp absolute summation, respectively. If the point source infinitely approaches the cusp, the magnifications of the three images will approach infinities, and the cusp relation $R_{\textrm{cusp}}$ will be close to 0.

In the middle panel of Fig. 1, a similar magnification relation holds when the source lies near a fold caustic. In this case, two images lie closely together, straddling the critical curve. One of two images is a minimum and the other one is a saddle. The fold image pair also has an asymptotic magnification relation \citep{bla86,sch92b,mao92,kee05,chu15}
\begin{equation}
R_{\textrm{fold}}=\frac{S_{\textrm{fold}}}{S_{|\textrm{fold}|}}= \frac{\mu_{\textrm{A}}+\mu_{\textrm{B}}}{|\mu_{\textrm{A}}|+|\mu_{\textrm{B}}|} ,
\end{equation}
where $\mu$ are the signed magnifications of the double images A and B. Here, $S_{\textrm{fold}}$ and $S_{|\textrm{fold}|}$ are fold summation and fold absolute summation, respectively. If the point source infinitely approaches the fold line, the magnifications of the two images will approach infinities, and the fold relation $R_{\textrm{fold}}$ will also be close to 0. In general, when a point source lies just on the cusp point or fold line, the numerators $S_{\textrm{cusp}}$ and $S_{\textrm{fold}}$ are usually not equal to 0 \citep{chu13,chu15}.

In many observed strong lenses, the positions of most multiple images can be fitted adequately using simple smooth lens models. Nevertheless, the observed flux ratios are more difficult to match \citep{koc91}. Actually, most of the observed fluxes of image pairs and triplets disagree with the fold and cusp relations. The discrepancy between the theoretical prediction and observed flux ratios is commonly referred to as the anomalous flux ratio problem \citep{mao04,con05,mck07,shi08}. Currently, the most favoured explanation of the flux ratio anomalies invokes the perturbation effects from small-scale structures hosted by lensing galaxies \citep{mao98,xu15}. Some studies suggest the consistency between the cold dark matter model and the observed flux ratios of the multiple images \citep{met02,chi02,bra04}, while some others find that subhaloes from cold dark matter simulations are insufficient to explain the observed flux-anomaly frequency in radio wavelengths \citep{ama06,mac06,che11,xu09,xu15}. Most of the previous works studied the anomalous flux ratio problem mainly based on cusp and fold types. In this work, we use much more precise numerical technique to study the magnification relations for the smooth SIE lens models, and also focus on the quad images of Einstein cross type which are seldom studied for the anomalous flux ratio problem.

This paper is arranged as follows. In Section 2, by dividing and ray-tracing in different image regions, we introduce a numerical method to calculate the positions and magnifications of different images for a point source in detail. In Section 3, based on these magnifications, we calculate the signed cusp and fold relations, and then discuss some observed cusp type lens examples. In Section 4, we define two new quantities which are mainly used for Einstein cross type, called cross relation and distance ratio. Then, we study the cross relations and distance ratios for nine observed Einstein crosses. Finally, in Section 5, the conclusions of this paper and a discussion are given.

\section{Calculation for image positions and magnifications}

\begin{figure}
\centering
\includegraphics[width=0.45\textwidth]{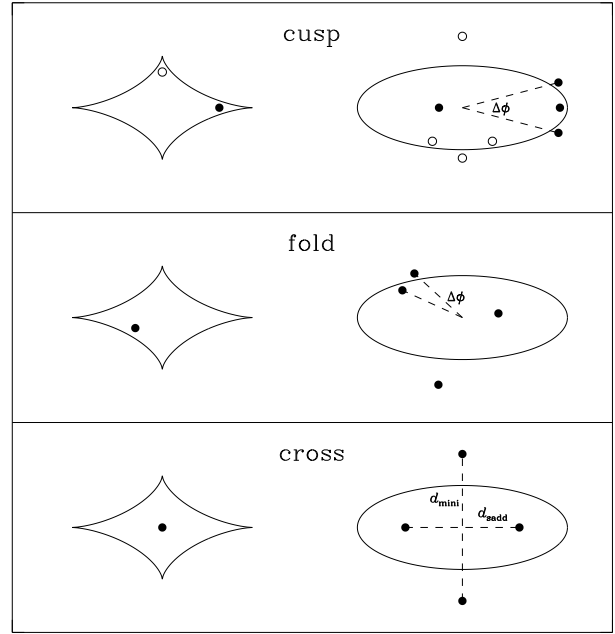}
\caption{Three basic configurations of quad lenses: cusp (top), fold (middle), and cross (bottom). In each panel, the figure on the left shows the caustic and source position in the source plane, while the figure on the right describes the critical curve and image positions in the lens plane. On the top panel, the solid and open circles show the major and minor cusp configurations, respectively.}
\end{figure}

The SIE lens can be derived by changing $\theta$ into $\sqrt{q^{2}x^{2}+y^{2}}$ through the singular isothermal sphere (SIS) lens. Here, $q$ is the axial ratio set for surface mass density or lens potential. Therefore, there are two ways to obtain it. One is to change $\theta$ in the convergence (or surface mass density in units of the critical surface mass density) $\kappa$, and it is singular isothermal elliptical density (SIED) lens \citep{kee98,kee00}
\begin{equation}
\kappa_{\textrm{SIED}}(x,y)=\frac{b}{2\sqrt{q^{2}x^{2}+y^{2}}} .
\end{equation}
Here, $b$ is used as a constant parameter which is equivalent to the Einstein radius. The other way is to change $\theta$ in the lens potential $\psi$, and it is singular isothermal elliptical potential (SIEP) lens \citep{kas93,kor94}
\begin{equation}
\psi_{\textrm{SIEP}}(x,y)=b\sqrt{q^{2}x^{2}+y^{2}} .
\end{equation}
Based on the two-dimensional Poisson equation $\nabla^{2}\psi=2\kappa$, the convergence $\kappa$ and lens potential $\psi$ can be calculated from each other. Then, the scaled deflection angles $\boldsymbol{\alpha}$ are analytically derived through the gradient equation $\boldsymbol{\alpha}=\nabla\psi$. The magnifications $\mu$ of the two lens models are all $\mu=1/(1-2\kappa)$.

Fig. 1 shows the three basic configurations of the quad images as well as caustic and critical curve of the SIED lens. The critical curve of the SIED lens is $q^{2}x^{2}+y^{2}=b^2$, which was derived by setting $1/\mu=0$. The left and right cusps on the caustic are major cusps, while the up and down ones are minor cusps. The radial critical curve of SIED lens degenerates into a point in the lens centre, so there is no Fermat maximum image there. The area of the astroid region (i.e., the region inside of the astroid caustic) decreases when the $q$ increases. For $q=1$, the lens returns to the SIS lens, and the area of the astroid region is 0. In this work, we mainly use the analytical SIED lens model and set $q=0.4$ through the paper. In addition, for numerical calculations, the grid of the lens plane is $1024\times1024$, while the grid of the source plane is $1000\times1000$. They correspond to the resolutions of 0.002 and 0.001 $\theta_{\textrm{E}}$ pixel$^{-1}$ approximately.

At first, we identify the subscript of each pixel in the astroid region. Secondly, we do reverse ray-tracing from the observer to each pixel of the lens plane, and then the light rays deflect to the source plane through the analytical deflection angles. In the lens plane, the pixels corresponding the light rays falling inside of the astroid caustic constitute the area of the image of the whole astroid region. The right panel of Fig. 2 shows the image of the astroid region, and as the numbers denote in Fig. 2, the image of the astroid region can be divided into four parts. Each point source lying in the astroid region has four images. Two minimum images lie in image regions 1 and 3, and their magnifications are positive. The other two saddle images lie in image regions 2 and 4, and their magnifications are negative. Here, the magnification values in the right panel of Fig. 2 are obtained from the left panel.

\begin{figure}
\centering
\includegraphics[width=0.45\textwidth]{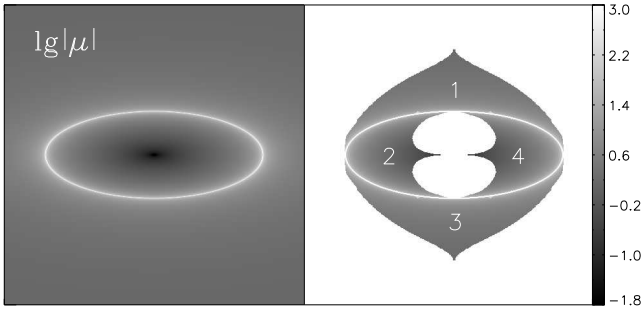}
\caption{The left panel shows the absolute value of the magnification $\mu$ in logarithm. The right panel lays out the four image regions corresponding to the astroid region, where the grey values are derived from the left panel.}
\end{figure}

Thirdly, to derive the magnifications of the four images for an arbitrary point source $\mathcal{P}$ inside the caustic region, the key point is to find the exact corresponding positions of the four images in lens plane. For example, in order to find the image position of $\mathcal{P}$ in image region 1, we simulate by shooting a bundle of light rays from the observer to each pixel of image region 1, and then the deflected light rays will land on the source plane. Among these light rays, there must be one light ray hitting the source plane nearest to the source $\mathcal{P}$, and the pixel corresponding to this light ray in image region 1 is the approximate image position of the source $\mathcal{P}$. This simple method is accurate to the pixel scale in the lens plane. Then, we set this pixel as an initial value, much more precise image position to any accuracy in image region 1 can be derived by iterative interpolations \citep{li05}. Similarly, we calculate the positions of the four images accurately in the four image regions for each pixel in the astroid region.

\begin{figure}
\centering
\includegraphics[width=0.45\textwidth]{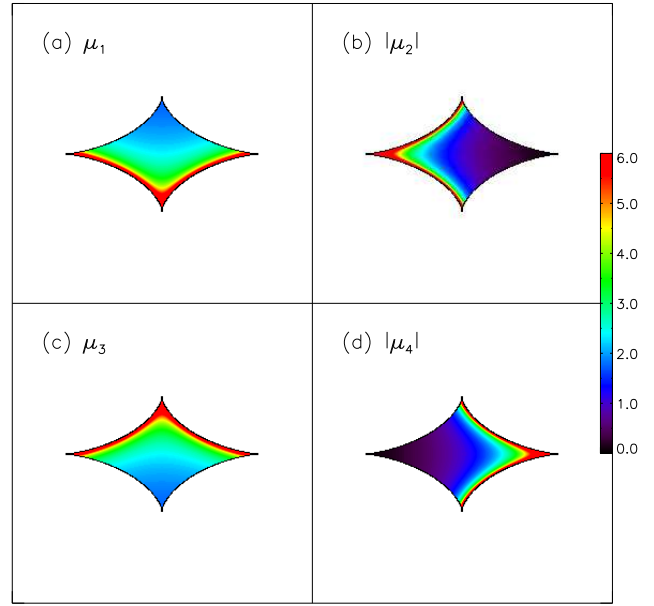}
\caption{The magnifications as functions of source positions. The values are mapped from four image regions to the source plane. Here, the saddle images $\mu_{2}$ and $\mu_{4}$ are shown in absolute values.}
\end{figure}

Finally, having derived the positions of the four images and also known the magnifications $\mu$ of the SIED lens for each position of the lens plane, we then map the magnifications of the four regions in the right panel of Fig. 2 on to the source plane accurately, and the results are shown in Fig. 3. For each panel, there are two fold lines near which the magnifications are infinities. One can also find that the smallest values of $\mu_{1}$ and $\mu_{3}$ must be larger than 1, which means the minimum images are never demagnified \citep{dal98}, while the smallest values of $|\mu_{2}|$ and $|\mu_{4}|$ only need to be larger than 0.

\section{The cusp and fold relations of SIE lenses}

\subsection{The generalized cusp and fold relations}

\begin{figure}
\centering
\includegraphics[width=0.45\textwidth]{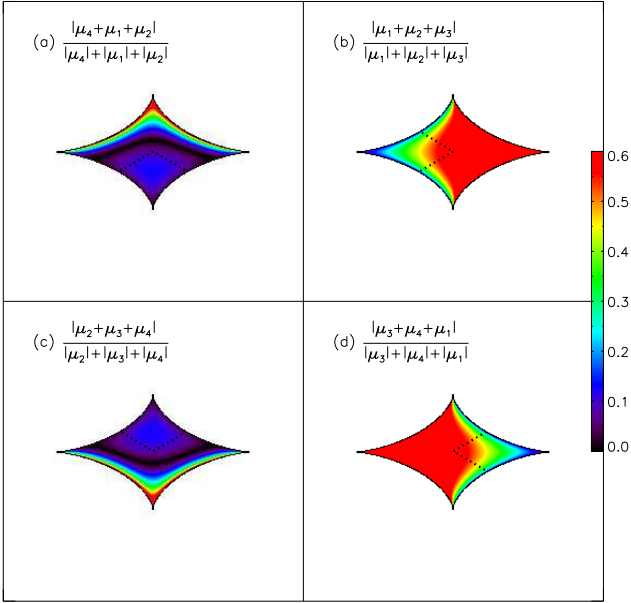}
\caption{The generalized cusp relations. The corner denoted by dotted lines in each panel means within which the three images calculated for cusp relation are the closest triplet among the four.}
\end{figure}

Based on these magnification maps shown in Fig. 3, we calculate the $R_{\textrm{cusp}}$ through equation (1), and the results are shown in Fig. 4. We call them generalized cusp relation for the reason that they show the values for all astroid region, although only the corners distinguished by dotted lines have physical meanings. In each panel of Fig. 4, the two fold lines connected to the dotted lines have the minimum value of absolute cusp relation 0, while the other two fold lines have the maximum value of absolute cusp relation 1. The generalized cusp relation maps are displayed in absolute values in Fig. 4, so one can find black stripes which outline the trends of $R_{\textrm{cusp}}=0$ curves inside of the astroid region in Fig. 4(a) and (c). The $R_{\textrm{cusp}}=0$ curves divide each of the two astroid regions into two parts. Below the $R_{\textrm{cusp}}=0$ curve of Fig. 4(a), the signs of $R_{\textrm{cusp}}$ are positive, while above this curve the signs are negative. In Fig. 4(c), the situation is opposite to that of Fig. 4(a). For convenience, we call them positive parts and negative parts based on the signs of $R_{\textrm{cusp}}$. However, the signed $R_{\textrm{cusp}}$ in Fig. 4(b) and (d) are all positive values. In the left two panels of Fig. 4, the proportions of the positive parts depend on axial ratio $q$. In this figure, the axial ratio of the SIED lens is still $q=0.4$. With increasing of the $q$, the areas of the positive parts in Fig. 4(a) and (c) decrease, and when $q\ga 0.68$, the area of the positive part is 0 \citep{chu15}.

Similarly, we also calculate the generalized fold relation $R_{\textrm{fold}}$ through equation (2), and they are shown in Fig. 5. In each panel, the fold line enclosed by the two dotted lines has the minimum value of absolute fold relation 0, while the other two neighbouring fold lines have the maximum value of absolute fold relation 1. The absolute fold relation of the remaining fold changes continuously on the fold line between 0 and 1. Similarly to Fig. 4(a) and (c), each panel can be divided into two parts by $R_{\textrm{fold}}=0$ curve which is also described by the black stripe. The signs of $R_{\textrm{fold}}$ on two sides of the $R_{\textrm{fold}}=0$ curve are different. In all panels of Fig. 5, the large parts have positive signs of fold relations, while the rest small parts have negative signs.

\begin{figure}
\centering
\includegraphics[width=0.45\textwidth]{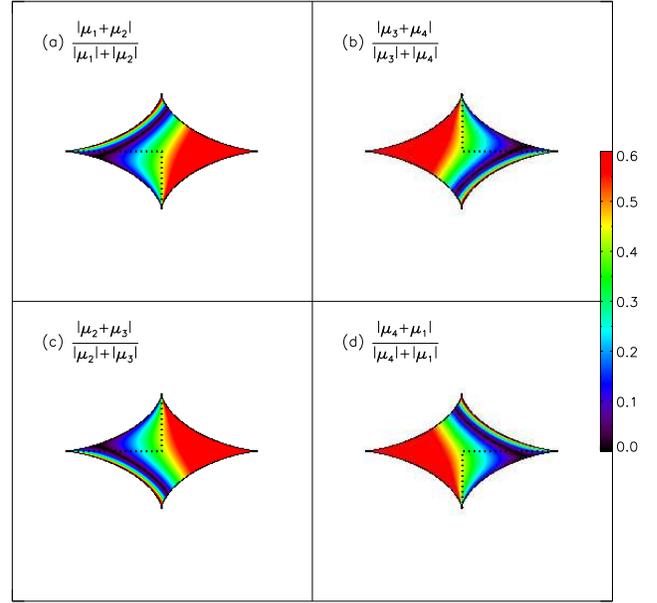}
\caption{The generalized fold relations. The quarter enclosed by dotted lines in each panel means within which the two images calculated for fold relation are the closest doublet among the four.}
\end{figure}

\begin{figure}
\centering
\includegraphics[width=0.45\textwidth]{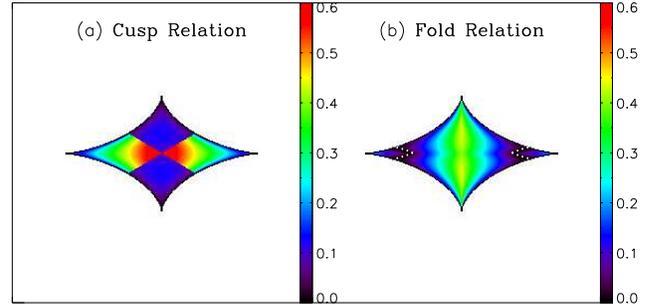}
\caption{The cusp and fold relations for SIED lens with $q=0.4$.}
\end{figure}

As mentioned before, for the generalized cusp relation map, only one quarter area for each panel in Fig. 4 has physical meanings. Therefore, we cut a corner from each panel and then constitute them into real cusp relation map. The two major cusp corners are cut from Fig. 4(b) and (d), while the two minor cusp corners are cut from Fig. 4(a) and (c). The two lines used here to cut the corner are parallel to the edges of the diamond constituted by the four cusps. Similarly, by cutting one quarter area from each panel in Fig. 5, one can get the fold relation map. The cusp and fold relation maps are shown in Fig. 6(a) and (b), respectively. In Fig. 6(b), the white dotted curves (where $R_{\textrm{fold}}=0$) divide the whole astroid region into two regions with positive and negative signs of $R_{\textrm{fold}}$, and hereafter we call them positive region and negative region to distinguish with the positive and negative parts in Figs. 4 and 5. The negative region includes two small areas near the two major cusps. Unlike the fold relation map, for SIED lens with $q=0.4$ in Fig. 6(a), the signed cusp relations are always larger than 0.

In Fig. 6, it is very interesting to study the cusp and fold relations on the caustic. One can easily find that the cusp relation has $R_{\textrm{cusp}}=0$, for point source lying on any position of the astroid caustic. However, for point source lying on the fold line and not very near to the cusps, the fold relation satisfies $R_{\textrm{fold}}=0$. The signed fold relations are $R_{\textrm{fold}}=\mp1/3$ on the major and minor cusps respectively \citep{kee05,aaz06}, because they correspond to $S_{\textrm{fold}}\to\mp\infty$ on the major and minor cusps respectively.

\subsection{The cusp and fold relations in observations}

In observations, for cusp type configuration, one can calculate the opening angle $\Delta\phi$ with respect to the lens centre for the three closest images. As shown in Fig. 1, in the cusp triple images, the opening angle $\Delta\phi$ is calculated based on the two outer images of the close triplet. Similarly, for fold type configuration, the opening angle $\Delta\phi$ of the two closest images can also be measured. Fig. 7(a) shows the distribution of the signed cusp relation as functions of the opening angle. The red points correspond to major cusp corners, while the green points correspond to minor cusp corners. Just as previous views \citep{kee03,ama06,xu09}, the smaller the opening angles, the smaller the cusp relations are. Fig. 7(b) shows the distribution of the signed fold relation as functions of the opening angle. We also confirm that the fold relation decreases with the opening angle decreasing. For the SIED lens, the fold relation shows a good linear relationship with respect to the opening angle. The points smaller than 0 in Fig.7(b) correspond to the negative region near the major cusps in Fig. 6(b). Additionally, the maximum $\Delta\phi$ of the triple and double images are $180^{\circ}$ and $90^{\circ}$ respectively, and they all correspond to the points lying in the centre of the source plane.

\begin{figure}
\centering
\includegraphics[width=0.5\textwidth]{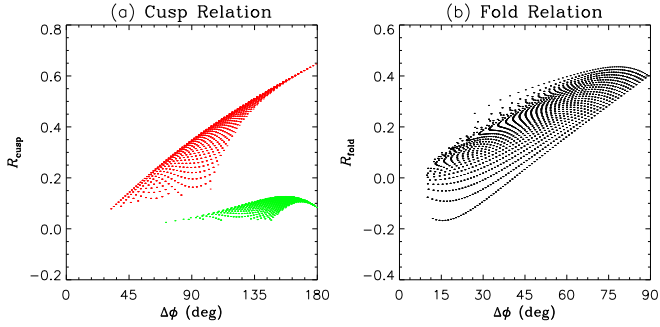}
\caption{The distributions of cusp relation (red for major cusps while green for minor ones) and fold relation of SIED lens as functions of the opening angles $\Delta\phi$.}
\end{figure}

In Fig. 6(b), the negative region near two major cusps is much smaller than the positive region, which has also been shown in the early work \citet{kee05}. Here, we should note that again the negative region is only relevant for the fold relation map or is considered for studying the fold image pairs. Unfortunately, the analytical solution of the white dotted curves is very difficult to be derived. Here, we calculate the areas of the positive and negative regions numerically by counting the pixel numbers with positive or negative signs in the astroid region in Fig. 6(b). The areas of the positive and negative regions for the SIEP lens are also derived similarly. Then, proportions of positive regions as functions of $q$ are calculated as the star and plus signs show in Fig. 8. We find that the larger the axial ratio $q$, the smaller the proportions of positive regions are. We speculate that when $q$ approaches to 0 or 1, the proportions of positive regions of the two SIE lenses will approach to 1 and 0.5 approximately.

\begin{figure}
\centering
\includegraphics[width=0.4\textwidth]{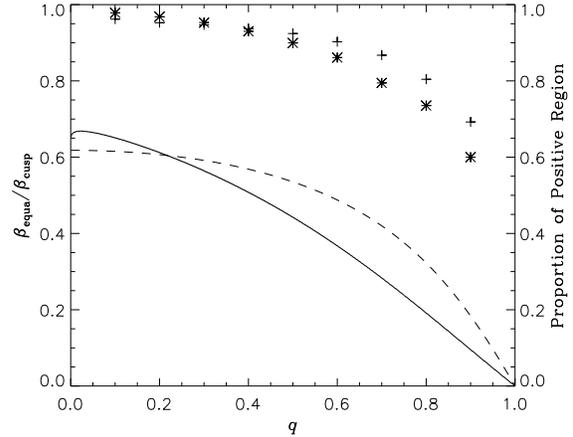}
\caption{The solid (SIED) and dashed (SIEP) curves show $\beta_{\textrm{equa}}/\beta_{\textrm{cusp}}$ in fold relation maps as functions of $q$, while the star (SIED) and plus (SIEP) signs show proportions of positive regions as functions of $q$, respectively.}
\end{figure}

In this situation, we also obtain another quantity $\beta_{\textrm{equa}}/\beta_{\textrm{cusp}}$ analytically, which reflects the proportion of positive region in another aspect. Here, $\beta_{\textrm{equa}}$ means the position on the major axis where the fold image pair have equal fluxes, while $\beta_{\textrm{cusp}}$ is the angular distance from the major cusp to the centre of the source plane, so larger $\beta_{\textrm{equa}}/\beta_{\textrm{cusp}}$ means larger proportion of positive region. Fig. 8 also shows the analytical results of $\beta_{\textrm{equa}}/\beta_{\textrm{cusp}}$ for the two SIE lenses. For larger $q$, proportion of the positive (negative) region of SIED lens is smaller (larger) than that of SIEP lens, while for very small $q$, the situation is contrary. When axial ratio $q$ approaches to 1, the two $\beta_{\textrm{equa}}/\beta_{\textrm{cusp}}$ tend to be 0. It is very interesting that when $q$ equals to 0, the $\beta_{\textrm{equa}}/\beta_{\textrm{cusp}}$ of the SIEP lens exactly equals to the Golden section $(\sqrt{5}-1)/2\approx0.618$.

For fold type lensed quasar, if the flux of the minimum is larger than that of the saddle, the source lies in positive region of the fold relation map, while if the flux of the saddle is larger than that of the minimum, it lies in negative region. In observations of many fold image pairs, the fluxes of the minimum images are usually larger than those of the saddle ones \citep{wit95,sch02}. Some works show that the existence of the substructures would suppress the flux of the saddle image and increase that of the minimum image \citep{koc04b,bra04}. However, in our high precise numerical work, the smooth SIE lenses could also bring this effect in statistics, since the negative regions are much smaller than the positive regions in the fold relation maps of the SIE lenses as shown in Fig. 6(b). Here, we are inclined to believe that the combination of the two effects is more probably.

In addition, the small negative region nearing the major cusps also means that in this small region the middle image has the largest flux among the major cusp triplet. If point source moves out of that small negative region and to the inward direction, the flux(es) of the side image(s) will exceed that of the middle image. This could explain that in some observed lens systems, not only do the cusp relations deviate from 0, but also the middle images are not the largest ones.

For cusp type lensed quasar RX\,J1131--1231 \citep{slu03}, the flux of the middle image is the largest one among the triple images, and the weighted average of cusp relation values for this sample is $0.355\pm0.015$ \citep{kee03}. For another cusp type example B1422+231 \citep{pat92} whose cusp relation is $0.187\pm0.004$ \citep{koo03}, the middle image also has the largest flux among the triplet \citep{nie14}. Without carrying out detailed lens modelling that includes local density perturbation effect, the fact that the saddle images have the largest flux among the triple images for the two systems can be explained if the sources lie in the negative regions of fold relation maps as shown in Fig. 6(b).

A typical flux-anomaly cusp type lens is B2045+265 \citep{fas99,mck07}, whose cusp relation is $0.501\pm0.035$ \citep{koo03}. The middle image has the smallest flux among the triplet, so B2045+265 was thought to strongly violate the cusp relation \citep{kee03,mck07}. Similarly, without considering the perturbation of the substructures, we suppose that the source quasar lies in the positive region of fold relation map. Then, we can naturally understand that the fluxes of two side images can surpass that of the middle image, and the flux ratios can be thought to be normal. In Fig. 6(b), the smaller the negative region is, the easier the fluxes of two side images surpass that of the middle one. Therefore, under the naive assumption that B2045+265 is a simple elliptical lens, as Fig. 8 shows, the projected mass distribution of this lens system prefers small axial ratio $q$.

In this subsection, the discussions on the three major cusp type lenses are only based on the positions and the fluxes of the images, and we suppose all of the images are not magnified by micro-lensing. In fact, the existence of the local density perturbations can change the shape of the caustic as well as the white dotted curves in Fig. 6(b), so the areas of the positive and negative regions will also be increased or decreased by the perturbations. Here, we do not study anomalous flux ratio problem for cusp or fold types in statistics, and only try to understand some specific instances.

\section{Cross relation and applications on Einstein crosses}

\subsection{Cross relations in some analytical lens models}

There are three basic configurations for quad or elliptical lenses, known as cusp, fold and cross types. Lots of works study the first two kinds of configurations mainly through the cusp relation and fold relation. Enlightened by the cusp and fold relations, we define a new quantity \emph{cross relation} by
\begin{equation}
R_{\textrm{cross}}=\frac{S_{\textrm{cross}}}{S_{|\textrm{cross}|}}= \frac{\mu_{\textrm{A}}+\mu_{\textrm{B}}+\mu_{\textrm{C}}+\mu_{\textrm{D}}} {|\mu_{\textrm{A}}|+|\mu_{\textrm{B}}|+|\mu_{\textrm{C}}|+|\mu_{\textrm{D}}|} .
\end{equation}
Here, $S_{\textrm{cross}}$ and $S_{|\textrm{cross}|}$ are \emph{cross summation} and \emph{cross absolute summation}, respectively. The cross relation $R_{\textrm{cross}}$ describes the magnification discrepancy between the minimum images and saddle images. For quad lenses, the two minimum images have positive magnifications, while the two saddle ones have negative magnifications. Therefore, larger cross relation represents larger differences of amplification abilities between minimum and saddle images.

\begin{figure}
\centering
\includegraphics[width=0.45\textwidth]{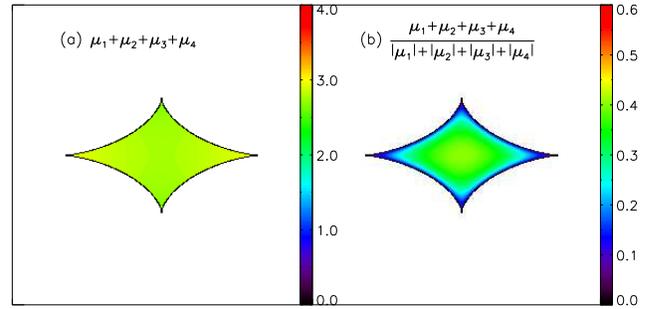}
\caption{The (a) generalized cross summation and (b) generalized cross relation for SIED lens with $q=0.4$.}
\end{figure}

Based on the four magnifications in Fig. 3, we calculate the $S_{\textrm{cross}}$ and $R_{\textrm{cross}}$ through equation (5), and the results are shown in Fig. 9. We call them generalized cross summation and relation, since they serve for any types of the three configurations. The cross summation of quad lens also describes magnification invariant. The magnification invariant means that, for some specific lens models, the sum of signed magnifications for all lensed images of a given point source is a constant, i.e., $I=\sum_{i}\mu_{i}$ \citep{dal98}. For the SIED and SIEP lenses, the magnification invariants are only valid when four images are produced. It is very interesting and surprising that the invariants are independent of most of the model parameters. The magnification invariant of SIEP lens is 2, and does not depend on the axial ratio $q$ \citep{dal98}. \citet{wit00} found the magnification invariant of SIED lens is approximately 2.8, which is nearly independent of the axial ratio $q$ and slightly depends on the position of the point source.

Based on the generalized cross summation in Fig. 9(a), we also check the magnification invariant of the SIED lens. For the SIED lens, our numerical result confirms that the cross summations (magnification invariants) near the major cusps are slightly larger than those near the minor cusps \citep{wit00}. On the other hand, the smooth and uniform colour distribution in Fig. 9(a) also prove that our method to calculate the magnifications is reliable. In this numerical work, we find the magnification invariant of SIED lens is $2.788\pm0.045$ for axial ratio $q=0.4$. We also test the magnification invariant of the SIEP lens through the same method, and the result also shows very good invariability about 2.

\begin{figure}
\centering
\includegraphics[width=0.4\textwidth]{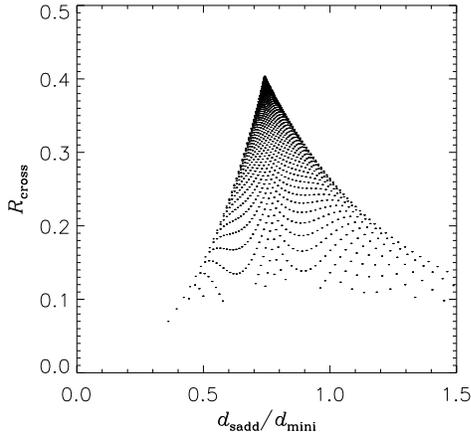}
\caption{The distributions of generalized cross relation of SIED lens as a function of the distance ratio for $q=0.4$.}
\end{figure}

In Fig. 9(b), for point sources infinitely near the caustic, $R_{\textrm{cross}}$ are close to 0 because of the infinite $S_{|\textrm{cross}|}$. In addition, one can easily find that the largest value of $R_{\textrm{cross}}$ always lies in the centre of the astroid region. We also calculate the cross relations for many other values of axial ratio $q$, and they all show the same. To manifest the distribution of Fig. 9(b), we display the cross relation $R_{\textrm{cross}}$ as a function of the distance ratio $d_{\textrm{sadd}}/d_{\textrm{mini}}$ in Fig. 10. Here, $d_{\textrm{sadd}}$ is angular distance between the two saddle images, while $d_{\textrm{mini}}$ is angular distance between the two minimum images as shown in Fig. 1. The black dots in Fig. 10 distribute like a triangle. The upper corner corresponds to the centre of the astroid region, while the left and right corners correspond to the minor and major cusps respectively.

For point source lying in the centre of source plane, the magnifications and positions of the four images of SIE lenses can be analytically calculated. Accordingly, the cross relation and distance ratio for Einstein cross type can also be easily derived. For the SIEP lens, they are
\begin{equation}
\begin{split}
R_{\textrm{cross}}=\frac{1-q^2}{1+q^2}=\epsilon , \quad  d_{\textrm{sadd}}/d_{\textrm{mini}}=q=\sqrt{\frac{1-\epsilon}{1+\epsilon}}  .
\end{split}
\end{equation}
For the SIED lens, it also has analytical results, whereas the functions are very complex. Therefore, we do not list them here. The cross relations and distance ratios of the two lens models are all independent of the Einstein radii. Fig. 11(a) and (b) show $R_{\textrm{cross}}$ and $d_{\textrm{sadd}}/d_{\textrm{mini}}$ of the two SIE lense as functions of the axial ratio $q$ for Einstein cross type.

As shown in Fig. 11(a), the $R_{\textrm{cross}}$ of SIE lenses decrease with increasing of axial ratio $q$. For the same $R_{\textrm{cross}}$, the $q$ of SIEP lens is larger than that of SIED lens. We can understand it like that, for any given ellipsoid lens, the contour ellipse of the lens potential is rounder than the contour ellipse of the mass density \citep{kas93}, and it means the potential contour has larger $q$ than the mass density contour. Therefore, also similarly in Fig. 11(b), having the same properties, the SIEP lens usually need larger $q$ compared to the SIED lens. In fact, the axial ratio $q$ for density and the axial ratio $q$ for potential are two different quantities, although they have the same name. It should be noted that, even for the SIED lens we can also derive the axial ratio $q$ for potential by fitting its potential contour, and vice versa.

We also calculate the cross relations and distance ratios for the other three lens models which are introduced in detail in our previous work \citep{chu15}. The two quantities of the three models also do not depend on the Einstein radii. For singular isothermal quadrupole (SIQ)\footnote{The lens potential is $\psi=b\theta-\gamma b \theta\cos 2\phi$. Here, parameter $\gamma$ describes of the strength of the quadrupole moment, and $\phi$ is the phase angle in the lens plane. This lens model is also called SIS+elliptical lens.} and SIS+shear\footnote{The lens potential is $\psi=b\theta-(\gamma/2)\theta^{2}\cos 2\phi$. Here, $\gamma$ is the external shear which do not contribute to external convergence, and $\phi$ is the phase angle.} lenses with point sources lying in the centres of the source planes, it is very interesting that, both the two lenses have
\begin{equation}
\begin{split}
R_{\textrm{cross}}=\gamma , \quad  d_{\textrm{sadd}}/d_{\textrm{mini}}=\frac{1-\gamma}{1+\gamma} .
\end{split}
\end{equation}
For Point+shear lens (also called Chang--Refsdal lens, \citealt{cha79,cha84}), the cross relation and the distance ratio for point source lying in the centre of source plane are similar to those of the SIEP lens
\begin{equation}
R_{\textrm{cross}}=\gamma , \quad  d_{\textrm{sadd}}/d_{\textrm{mini}}=\sqrt{\frac{1-\gamma}{1+\gamma}}  .
\end{equation}

Coming back to Fig. 10, one can find that most black points lie near the upper corner. Taking into account that these points are uniformly sampled in Fig. 9(b), we can conclude that if the point source slightly moves away from the centre of source plane, the cross relation and distance ratio do not change significantly. Therefore in observations, the slight deviations from the perfect Einstein crosses will not affect the two quantities too much. Using the observed cross relations or distance ratios of quad lens systems, in theory, through Fig. 11(a) and (b) we can give the constraints on the axial ratios $q$ under the assumptions that these lenses are described by SIED or SIEP models. In practice, it is very difficult to constrain them only using the cross relations, because the uncertainties in image fluxes are much larger than those in image positions. Moreover, to judge which lens model is more real is also not easy to carry out.

\begin{figure}
\centering
\includegraphics[width=0.46\textwidth]{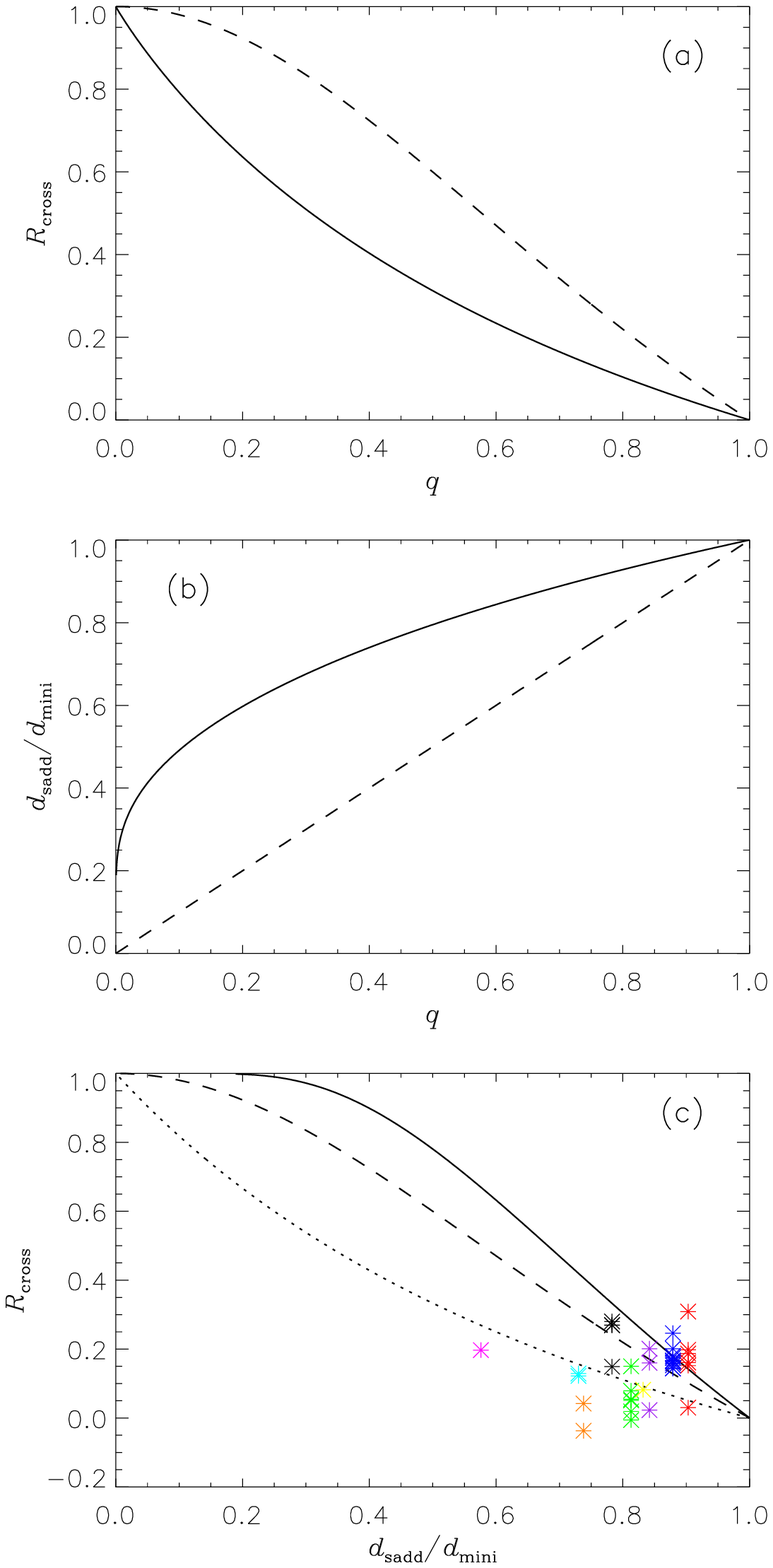}
\caption{(a) The solid (SIED) and dashed (SIEP) curves show the cross relations as functions of $q$ for point sources lying in the centres of the source planes. (b) Similar to (a), but for distance ratios. (c) The solid (SIED), dashed (SIEP and Point+shear) and dotted (SIQ and SIS+shear) curves show the cross relations as functions of distance ratios. The nine different coloured stars represent nine lens samples of Einstein cross type (Q2237+0305: red, H1413+117: green, HE\,0435--1223: blue, SDSS\,1138+0314: yellow, HST\,14176+5226: cyan, HST\,12531--2914: brown, HST\,14113+5211: magenta, J1011+0143: violet, J1149+2223: black). }
\end{figure}

\subsection{Cross relations in observed Einstein crosses}

\begin{table*}
 \centering
 \caption{Cross relations and distance ratios for nine quad lenses with Einstein cross configuration.}
 \begin{tabular}{@{}lcccc}
  \hline

  Lens & $d_{\textrm{sadd}}/d_{\textrm{mini}}$ & $R_{\textrm{cross}}$ & Band & Reference \\
  \hline
  Q2237+0305      & $0.903\pm0.003$  & $0.152\pm0.053$ & IR         & \citet{ago00}  \\
                  &                  & $0.030\pm0.053$ & IR         & \citet{tro02}  \\
                  &                  & $0.198\pm0.109$ & \emph{V}   & \citet{dai03}  \\
                  &                  & $0.309\pm0.055$ & X-ray      & \citet{dai03}  \\
                  &                  & $0.161\pm0.023$ & mid-IR     & \citet{min09}  \\
                  &                  &   0.187         & UV/optical & \citet{ass11}  \\
  H1413+117       & $0.813\pm0.003$  & $0.019\pm0.017$ & \emph{H}   & \citet{kne98} \\
                  &                  & $0.051\pm0.052$ & optical/UV & \citet{cha99} \\
                  &                  & $0.054\pm0.039$ & mid-IR     & \citet{mac09} \\
                  &                 & $-0.006\pm0.003$ & \emph{r}   & \citet{goi10} \\
                  &                  & $0.150\pm0.034$ &            & \citet{hut10} \\
                  &                  &   0.079         & UV/optical & \citet{ass11}  \\
  HE\,0435--1223   & $0.879\pm0.002$  &      0.144      & \emph{g}  &  \citet{wis02}  \\
                  &                  &      0.153      & \emph{r}  &  \citet{wis02}  \\
                  &                  &      0.164      & \emph{i}  &  \citet{wis02}  \\
                  &                  & $0.167\pm0.007$ & \emph{R}  &  \citet{ric11}  \\
                  &                  & $0.159\pm0.010$ & \emph{V}  &  \citet{ric11}  \\
                  &                  & $0.144\pm0.032$ & \emph{K}  &  \citet{fad11}  \\
                  &                  & $0.201\pm0.041$ &  $L'$     &  \citet{fad11}  \\
                  &                  & $0.180\pm0.006$ &  F160W    &  \citet{cou11}  \\
                  &                  & $0.175\pm0.011$ & \emph{H}  &  \citet{bra14}  \\
                  &                  & $0.246\pm0.038$ & Radio     &  \citet{jac15}  \\
  SDSS\,1138+0314  & $0.832\pm0.003$  &   0.082         & UV/optical & \citet{ass11} \\
  HST\,14176+5226  & $0.730\pm0.012$  & $0.129\pm0.036$ & \emph{V}  & \citet{rat95}  \\
                  &                  & $0.122\pm0.059$ & \emph{I}  & \citet{rat95}  \\
  HST\,12531--2914 & $0.738\pm0.032$  & $0.042\pm0.087$ & \emph{V}  & \citet{rat95}  \\
                  &                 & $-0.037\pm0.160$ & \emph{I}  & \citet{rat95}  \\
  HST\,14113+5211  & $0.576\pm0.016$  & $0.197\pm0.020$ & F702W     &  \citet{fis98} \\
  J1011+0143      & $0.842\pm0.014$  & $0.023\pm0.023$ & \emph{B}  & \citet{bol06}  \\
                  &                  & $0.201\pm0.064$ & F555W  & \citet{shu16a}  \\
                  &                  & $0.160\pm0.067$ & F814W  & \citet{shu16a}  \\
  J1149+2223 (SN Refsdal)  & 0.783   & $0.269\pm0.058$ & F140W  & \citet{kel15}  \\
                  &                  & $0.280\pm0.075$ & F125W  & \citet{kel15}  \\
                  &                  & $0.270\pm0.119$ & F105W  & \citet{kel15}  \\
                  &                  & $0.149\pm0.019$ &        & \citet{rod16}  \\
  \hline
 \end{tabular}

 \medskip
   All of the lensed sources are quasars, except a Ly$\alpha$-emitting galaxy in lens system SDSS\,J1011+0143 and SN Refsdal behind the galaxy cluster MACS\,J1149.6+2223. The image positions of the quasars came from the CASTLES survey, while flux (ratio) or magnitude information of the cross images are derived from the related references.
\end{table*}

In Table 1, we list nine quad lenses of Einstein cross type, including seven quasars and a galaxy and a supernova (SN). The seven quasar samples can be found in the CASTLES data base\footnote{http://www.cfa.harvard.edu/castles/} \citep{fal01}. Sometimes, Einstein cross is the peculiar name for Q2237+0305 which was discovered by \citet{huc85}. In addition, the cross lens H1413+117 discovered by \citet{mag88} also has its unique name Cloverleaf quasar. Here, we call all of the cross samples Einstein cross. In Table 1, the distance ratios are calculated through the position information of the four lensed images, while the cross relations are calculated based on the flux (ratio) values or magnitudes of the lensed images. Unfortunately, for the flux ratios (calculated from three images to the rest specific ones) of many lens samples, the divergences among different observation bands or among different references are usually very large.

In Fig. 11(c), the three curves based on equations (6)-(8) show cross relations $R_{\textrm{cross}}$ as functions of the distance ratios for five lens models. Each of theoretical cross relations is inversely correlated with the distance ratio. One can easily find that the $R_{\textrm{cross}}$ of the nine observed samples in fact do not fit these quad lens models very well. We speculate there are a few reasons which are as follows.
\begin{enumerate}
  \item  Due to electromagnetic (non-gravitational) effects such as extinctions by dusts or scattering by hot gases along different lines of sight \citep{kee03}, the fluxes of different images are affected depending on the wavelengths. Furthermore, estimations of the intrinsic fluxes of the lensed images can also be biased by foreground light subtraction method as indicated in the discussions of SDSS\,J1011+0143.
  \item \label{item2} The time delays among different images usually cover from few days to few years (summarized from \citealt{par10,rat15}). Since the majority of quasars exhibit continuum variability of the order of 20 per cent on time-scales of months to years \citep{hoo94,van04,wil08}, the time delays among different images can affect the real cross relation.
  \item  The substructures in the dark matter halo may play important role in the anomalous flux ratio problem \citep{chi02,met02,koc04b}. In principle, the substructures could also affect the cross relation and distance ratio at the same time.
  \item  The lens models are too simple to describe the real lenses both in radial and tangential profiles, and they should include high-order moment distributions and may exist external shear as well as other disturbances.
  \item  Some samples are not perfect Einstein crosses, because the sources are hardly lying on the optimum positions. This deviation effect usually decreases the cross relation as shown in Fig. 10.
  \item  There are micro-lensing effects in some observed images. The compact foreground lens object can change the flux of one image alone without perturbing the other ones \citep{sch10}. For detailed discussion about the micro-lensing effects of these Einstein crosses, please see the following paragraphs.

\end{enumerate}

Einstein cross Q2237+0305 is the first one that was reported to be a micro-lensing event in 1989 by \citet{irw89}. Since its discovery, this system has permanently shown uncorrelated fluctuations between the images \citep{kay89,wam90,ost96,woz00}. The two fluctuated micro-lensing images A and B are just the two Fermat minimum images \citep{koc04a,eig08}, which can significantly affect the cross relation. As shown by red stars in Fig. 11(c), Q2237+0305 has the largest distance ratio among the nine samples, and it also has very large scatter of cross relations among different bands. We suppose these large deviations from the theoretical cross relations are related to the miro-lensing effect.

Micro-lensing perturbations to the flux ratios of gravitationally lensed quasar images can vary with wavelength because of the chromatic dependence of the accretion disk's apparent size. Therefore, blue light from the inner regions is more strongly micro-lensed than red light from farther out \citep{wam91,bla11}. Actually, in IR band micro-lensing events are not observed for Q2237+0305 \citep{ago00,tro02}. \citet{ago00} considered that the much extended IR emission regions are unlikely to be micro-lensed as the point sources. Thus, the IR fluxes should measure the macro-magnification. Therefore, in Fig. 11(c) the cross relations calculated in IR band fit the theoretical curves better than those calculated in the other bands.

Cloverleaf H1413+117 was also considered to have micro-lensing effect in saddle image D which was significantly magnified \citep{ang90,hut93,ost97,ang08,hut10}. The cross relations of the Cloverleaf are shown by green stars in Fig. 11(c). However, \citet{cha04} found that the saddle image A was also enhanced by micro-lensing in the X-ray band. Nevertheless, each point of view can decrease the cross relation of this sample.

HE\,0435--1223 is an almost textbook example for gravitational lensing, with its four nearly identical components arranged symmetrically around a luminous early-type galaxy \citep{wis02}. The cross relations are shown in blue colour in Fig. 11(c). Micro-lensing was detected in a subsequent monitoring campaign \citep{wis03,koc06} which probably affects the minimum image A most strongly \citep{ric11,cou11,bra14}. Since the minimum image has positive sign of magnification, the magnified A component by micro-lensing effect also equals to increasing the cross relation.

Einstein cross lens system SDSS\,J1011+0143 was discovered by \citet{bol06}, who found the lens galaxy is a bright elliptical at $z_{\textrm{lens}}=0.331$, while the lensed source is a Ly$\alpha$-bright, star-forming galaxy at redshift $z_{\textrm{source}}=2.701$ from Keck \emph{B}-band image. The four lensed images form a perfect Einstein cross, and the cross relation inferred from Keck \emph{B}-band image has a relatively lower value. Later, based on higher resolution observations through the \emph{HST} F555W (\emph{V}) and F814W (\emph{I}) filters of the Wide Field Channel (WFC), \citet{shu16a} found the lens in fact comprises a merging galaxy pair with a small projected separation of $\approx 4.2$ kpc. As shown in Table 1, the cross relations derived from the \emph{HST} \emph{V}- and \emph{I}-band data are significantly larger than that derived from Keck \emph{B}-band data. The deviations stem from the differences in the relative brightness of image C. Normalized by the other three images, image C appears much more bright in the \emph{HST} data than that in the Keck data. The reasons for such differences are still uncertain. We suspect that they can be caused by the different foreground-light subtraction approaches. Unable to resolve the lens galaxy pair, \citet{bol06} used a radial b-spline model with a quadrupole angular dependence to model the light distribution of the foreground lens. Instead, \citet{shu16a} used four S\'{e}rsic components to model the two lens galaxies as well as the extra light presumably from the stripped materials. As image C is the minimum image with positive magnification, the \emph{HST}-derived cross relations are hence larger.

The four-image lensed SN Refsdal behind the MACS\,J1149.6+2223 cluster was discovered in \emph{HST} WFC3-IR image by \citet{kel15}. SN Refsdal is the first strongly lensed SN resolved into multiple images. The fluxes of the four images we used here are observed by \emph{HST} WFC3-IR photometry in 2014 November \citep{kel15}. Since SNe have much larger light curve changes than quasars, and the time delays among the maxima of the light curves of the four images can not be neglected, as also stated in item \ref{item2}, there is no doubt that the fluxes of the four images observed only in 2014 November will bring significant errors in the cross relation. Fortunately, by using a set of light curve templates constructed from the family of SN 1987A-like peculiar Type II SNe, \citet{rod16} calculated time delays and magnification ratios of SN Refsdal, and found the magnifications relative to image S1 are $1.15\pm0.05$ (S2), $1.01\pm0.04$ (S3), and $0.34\pm0.02$ (S4) respectively. The differences between the cross relations calculated based on the two works are very large as shown in Table 1 and Fig. 11(c). There is no doubt that the cross relation calculated by fitted magnifications through light curve templates is more reliable than those calculated through the fluxes observed at the same time.

Doing not take into account the obvious micro-lensing samples and the cross relations of SN Refsdal calculated from the fluxes observed at the same time, only through the observed data, Fig. 11(c) shows the lowest dotted curve fits them comparatively best, i.e., SIQ or SIS+shear lens model. However, the SIQ and SIS+shear lens models are apparently not the similar kinds, so this curve may include some other families of lens models. Furthermore, considering limitation of the sample size and the large discrepancies of cross relations among different observation bands, we can \emph{not} get the conclusion that the data approve the SIQ and SIS+shear models more, and we need other information to give more constraints. In addition, although the cross relations of the nine samples do not agree with the theoretical curves of quad lens models very well, they do not give obvious evidence for anomalous flux ratio as cusp and fold lens systems do. One possible explanation is that local density fluctuations produced by dark matter halo substructures are much less likely to perturb image fluxes of Einstein cross configuration than those of fold or cusp configurations, because the images of Einstein crosses are farther away from the critical curves than cusp or fold images.

\section{Conclusions and Discussion}

The four-image lens systems are very important and are very common in the observations of lensed quasars or galaxies \citep{rus01,cla02}. In this work, we mainly study the magnification relations for three basic configurations of quad lens models using numerical method. At first, by dividing and ray-tracing in different image regions, we provide a numerical method to determine the positions of different images of a point source precisely. Then, the magnifications of the four images can be derived consequently. Based on these magnifications, we calculate the cusp relation $R_{\textrm{cusp}}$ and fold relation $R_{\textrm{fold}}$ as functions of opening angle $\Delta\phi$. Being consistent with the previous views \citep{kee03,ama06,xu09}, our results also confirm that the smaller the opening angle, the smaller the two relations are. For the SIED lens, the fold relation $R_{\textrm{fold}}$ shows a good linear relationship with respect to the opening angle.

Through the magnifications in Fig. 3, we calculate magnification invariant which means the sum of signed magnifications for all lensed images of a given point source is a constant. The numerical results show very good invariants for SIED and SIEP lenses with different axial ratios $q$. On the other hand, it also proves our method to calculate the magnifications of the four images is reliable. The fold relation map in Fig. 6(b) can be divided into positive and negative regions which are defined according as the signs of the fold relations, and the positive region is significantly larger than the negative region. With increasing axial ratio $q$, the proportions of the negative regions increase. For SIED and SIEP lenses with same $q$ (for real projected dark matter halo lenses, $q$ can not be very small), the proportion of the positive (negative) region of SIED lens is usually smaller (larger) than that of SIEP lens.

In observations of many fold image pairs, the fluxes of the minimum images are usually larger than those of the saddle images \citep{wit95,sch02}. In this numerical work, by studying the proportions of the negative and positive regions, we find that this can be explained not only by the effect of substructures \citep{koc04b,bra04} but also by proper smooth SIE models with the point source lies in the positive region of the fold relation map. We prefer that the combination of the both is more probably. Then, combining the signed fold relation map, we discuss three lensed quasars of major cusp type, and give some constraints on the positions of the quasar sources as well as the ellipticities of projected lens haloes.

There are three basic configurations for quad or elliptical lenses, known as cusp, fold and cross types. Many works mainly use the cusp relation and fold relation to study the first two kinds of configurations. Enlightened by cusp and fold relations, we define a new quantity cross relation $R_{\textrm{cross}}$ which describes the magnification discrepancy between the minimum images and saddle images, as well as the quantity distance ratio $d_{\textrm{mini}}/d_{\textrm{sadd}}$. We derive the analytical cross relations $R_{\textrm{cross}}$ as functions of the distance ratios for five often used quad lens models with Einstein cross type. All of them do not depend on the Einstein radii. We also find that the SIEP and Point+shear models have the same relationship between the cross relation and distance ratio, and the SIQ and SIS+shear models also share the same one.

We calculate cross relations $R_{\textrm{cross}}$ and distance ratios for nine observed Einstein cross samples. In theory, for most of the quad lens models, the $R_{\textrm{cross}}$ decrease with increasing distance ratio. However, the $R_{\textrm{cross}}$ of the nine observed samples do not fit the quad lens models very well, and we propose several reasons for it. Although the observed data fit the SIQ and SIS+shear lens models comparatively better, considering the model family degeneracy and the limitation of the sample size as well as the large discrepancies of cross relations among different observation bands, we think we can \emph{not} get the conclusion that the data approve the two models more, and we need other information and more samples to give more stringent constraints. Furthermore, with large divergences among different observation bands, the cross relations of the nine Einstein crosses do not show obvious evidence for anomalous flux ratio.

In the future work, we can study the cross relations and distance ratios of elliptical lenses for the NFW \citep{nav97} radial profile by numerical method. Consequently, one can derive the two quantities as functions of the axial ratio $q$ or ellipticity more reliably. In the future, more and more perfect cross type quad lenses will be observed from large surveys or dedicated programmes (e.g., \citealt{ogu10,ser14,col15,shu16b}). One can get more confirmed information through the Einstein cross type than the other types of quad lenses because of the symmetry, such as that in the circular lenses, Einstein rings can tell us more confirmed information than two-image lenses. As we know, Einstein cross has the best symmetry among all quad lenses. Many observed cross images break the symmetry such as the two minimum (saddle) images having different fluxes, time delays, or distances from lens centre. These unsymmetry information can also be used to study the substructure or high-order distributions of lens body. Therefore, it is necessary to build some reliable models especially for the cross type lensed quasars. The existence of anomalous flux ratio in Einstein cross type can also be well tested through more precise observations and better lens models in the future.

\section*{Acknowledgements}

The authors are grateful to the referee for constructive comments and detailed suggestions to improve this manuscript, and also thank Yiping Shu and Qianli Xia for helpful discussion and valuable suggestions. We acknowledge the supports by the National Key Basic Research Program of China (2015CB857000) and the ``Strategic Priority Research Program the Emergence of Cosmological Structures" of the Chinese Academy of Sciences (XDB09000000). ZC is supported by National Natural Science Foundation of China (11403103) and China Postdoctoral Science Foundation (2014M551681). GL is supported by the One-Hundred-Talent fellowships of CAS and by the NSFC grants (11273061 and 11333008). WPL acknowledges supports by the NSFC projects (11473053, 11121062, 11233005, U1331201).

\label{lastpage}

\end{document}